\documentclass{PoS}
\let\ifpdf\undefined

\usepackage{graphicx}
\usepackage{multirow}
\usepackage[nice]{nicefrac}

\newcommand{\mc}[3]{\multicolumn{#1}{#2}{#3}}
\newcommand{\bi}{\begin{itemize}}
\newcommand{\ei}{\end{itemize}}

\newcommand{\be}{\begin{equation}}
\newcommand{\ee}{\end{equation}}
\newcommand{\bea}{\begin{eqnarray}}
\newcommand{\eea}{\end{eqnarray}}
\newcommand{\op}[1]{\ensuremath{\langle B_q^0 | \mathcal{O}_{#1} | \bar{B}_q^0 \rangle (\mu)}}
\newcommand{\me}[1]{\ensuremath{\langle B_q^0 | \mathcal{O}_{#1} | \bar{B}_q^0 \rangle}}
\newcommand{\shortop}[1]{\ensuremath{\langle{\cal O}_{#1}\rangle}}

\newcommand{\bt}[1]{\begin{table}[h!]\begin{tabular}{#1} \hline\hline  \\[-0.5em]}
\newcommand{\et}[2]{\hline\hline \end{tabular} \caption{#1} \label{#2} \end{table}}

\title{Neutral $B$ mixing from $2+1$ flavor lattice QCD:  the Standard Model and beyond}
\ShortTitle{$B_d^0$ and $B^0_s$ mixing}    

\author{
C.M.~Bouchard$^{\, *\ a,b,c}$,
E.D.~Freeland$^{\, *\ a,d}$,
C.~Bernard$^{\ e}$,
A.X.~El-Khadra$^{\ a}$,
E.~G\'amiz$^{\ b,f}$,
A.S.~Kronfeld$^{\ b}$,
J.~Laiho$^{\ g}$, and
R.S.~Van~de~Water$^{\ h}$
\hphantom{\speaker{C.M.~Bouchard and E.D.~Freeland}}
\\ \\
\llap{$^a$}Physics Department, University of Illinois, Urbana, IL  61801, USA \\
\llap{$^b$}Theoretical Physics Department, Fermi National Accelerator Laboratory,~Batavia, IL  60510, USA\\
\llap{$^c$}Department of Physics, The Ohio State University, Columbus, OH 43210, USA\\
\llap{$^d$}Department of Physics, Benedictine University, Lisle, IL 60532, USA\\
\llap{$^e$}Department of Physics, Washington University, St. Louis, MO 63130, USA\\
\llap{$^f$}CAFPE y Departamento de Fisica Teorica y del Cosmos, Universidad de Granada,
E-18002 Granada, Spain\\
\llap{$^g$}SUPA, School of Physics and Astronomy, University of Glasgow, Glasgow, G12 8QQ, UK\\
\llap{$^h$}Physics Department, Brookhaven National Laboratory, Upton, NY 11973, USA}
\author{Fermilab Lattice and MILC Collaborations\\
\email{bouchard.18@osu.edu} \textrm{ and } \email{eliz@fnal.gov}}

\abstract{
We report on the status of our lattice-QCD calculation of the hadronic contribution to $B_d^0$ and $B^0_s$ mixing, with $2+1$ flavors of dynamical sea quarks.
Preliminary results for hadronic mixing matrix elements are given for a basis of five four-quark, dimension-six, $\Delta B=2$ mixing operators that spans 
the space of all possible hadronic mixing contributions in the Standard Model and beyond.
At the intermediate stage of analysis reported on in this work, our errors are competitive with published Standard Model matrix element results.
For beyond the Standard Model matrix elements, this is the first unquenched calculation and the first new lattice-QCD calculation in ten years.}

\FullConference{ The XXIX International Symposium on Lattice Field Theory - Lattice 2011\\
July 10-16, 2011\\
Squaw Valley, Lake Tahoe, California}

\begin{document}

\section{Introduction}		\label{sec-intro}  

The main goal of particle physics today is the search for new physics. 
A primary focus of that search is the precision determination of the  Cabibbo-Kobayashi-Maskawa (CKM) matrix elements
or, more accurately, an overdetermination. 
To achieve this, many flavor-changing processes are studied by both experiment and theory and the results are combined to look for inconsistencies, which could indicate new physics at work~\cite{Laiho:2009eu, Lunghi:2009ke, Laiho:2011nz}.
At the same time, as hints of new physics appear, 
beyond the Standard Model phenomenologists use the information to constrain, support, or rule out new models~\cite{Lenz:2010gu, Bona:2008jn, BSM}.
In both cases, lattice-QCD calculations of hadronic matrix elements are crucial.

Neutral $B$ mixing is a promising place to look for new physics for two reasons.
First, in the Standard Model it occurs via a loop process in which the contributions are suppressed both by CKM matrix elements, and, for all but the top quark, masses of the loop quark.
So, any new physics could present a relatively large signal. 
Second, many recent reviews of the persistent 2--3$\sigma$ tension between the Standard Model and flavor physics experiments indicate that new physics in $B$ mixing is a likely explanation~\cite{Laiho:2009eu, Lunghi:2009ke, Laiho:2011nz, Lenz:2010gu, Bona:2008jn}.

The most general $\Delta B = 2$ effective hamiltonian contains eight dimension-six operators
\be
	\mathcal{H}_{\rm eff} = \sum_{i=1}^5 C_i \mathcal{O}_i  + \sum_{i=1}^3 \tilde{C}_i \tilde\mathcal{O}_i ,  \label{eq:hamiltonian}
\ee
with
\be
	\begin{array}{l l}
	\mathcal{O}_1 = (\bar{b}^\alpha \gamma_\mu L q^\alpha )  \; (\bar{b}^\beta \gamma_\mu L q^\beta ),	&
	\mathcal{O}_4 = (\bar{b}^\alpha L q^\alpha )  \; (\bar{b}^\beta R q^\beta ), \\
	\mathcal{O}_2 = (\bar{b}^\alpha L q^\alpha )  \; (\bar{b}^\beta L q^\beta ),	&
	\mathcal{O}_5 = (\bar{b}^\alpha L q^\beta )  \; (\bar{b}^\beta R q^\alpha ),	\\
	\mathcal{O}_3 = (\bar{b}^\alpha L q^\beta )  \; (\bar{b}^\beta L q^\alpha ),
	\end{array}  \label{eq:susybasis}
\ee
where $\alpha$ and $\beta$ are color indices, $L$ and $R$ are the projection operators $\frac{1}{2}( 1 \pm \gamma_5)$, and $\tilde\mathcal{O}_{1,2,3}$ are obtained from $\mathcal{O}_{1,2,3}$ by $L \to R$.

Parity conservation of QCD ensures the matrix elements for $\tilde\mathcal{O}_{1,2,3}$ are physically equivalent to those for $\mathcal{O}_{1,2,3}$, leaving five independent matrix elements.
The matrix  element \me{1} appears in the Standard Model expression for the $B$ meson mass difference $\Delta M_q$.
The matrix element \me{2} is needed for renormalization of \me{1}, and all three can be useful in studies of the width difference $ \Delta \Gamma$ and its constraints on new physics~\cite{Lenz:2006hd, CDF9787_D05928_Jpsiphi}.
The matrix elements \me{4} and \me{5} are needed for generic extensions beyond the Standard Model (some examples are given in Refs.~\cite{BSM}).

In the Standard Model,  
\be
	\Delta M_q = 
		\left( \frac{G_F^2 M_W^2 S_0}{4 \pi^2 M_{B_q}} \right)  \eta_B(\mu)  
		|V_{tb} V_{tq}^*|^2  \op{1},  \label{eq:massdiff}
\ee
where the quantities in parentheses and $\eta_B$ are known factors, $V_{ij}$ is a CKM matrix element, and we use relativistic normalization for the states.
Measurements of $\Delta M_q$ have sub-percent errors~\cite{Abulencia:2006ze, Nakamura:2010zzi},
so our ability to constrain the CKM matrix contribution $|V_{tb} V_{tq}^*|$ is limited by how precisely we know \me{1}.
Current, published lattice-QCD calculations~\cite{Gamiz:2009ku} report errors of just under 14\% on \me{1}.

For $B_s$ mixing only, an unquenched calculation of \me{1,2,3} appeared in Ref.~\cite{HPQCD:O123}, though it includes data at only one lattice spacing, $a \approx$~0.12~fm.
The only calculation to date of \me4 and \me5 appeared a decade ago~\cite{Becirevic:2001xt}.  In that work all five matrix elements were calculated using the quenched approximation for the sea, static heavy valence quarks with an interpolation to $m_b$, and a linear extrapolation in the light valence mass to reach $B_d$.
These approximations are no longer required in modern lattice-QCD calculations.  
We include three sea quarks to simulate up, down, and strange; work directly at $m_b$; include data at two or more lattice spacings; and use rooted, staggered chiral perturbation theory to guide a simultaneous chiral-continuum extrapolation.  This work presents the first full error budget for the beyond the Standard Model matrix elements.

Historically, it has been useful to parametrize the matrix elements as~\cite{Lenz:2006hd, Beneke:1996gn}
\be
	\op{i} = \mathfrak{c}_i\ M_{B_q}^2\ f_{B_q}^2\ B_{B_q}^{(i)}(\mu),  \label{eq:bagparams}
\ee
with coefficients $\mathfrak{c}_i=(\nicefrac{2}{3},\ \nicefrac{-5}{12},\ \nicefrac{1}{12},\ \nicefrac{1}{2},\ \nicefrac{1}{6} )$.   The $B_{B_q}^{(i)}$, known as bag parameters, characterize deviation from the so-called vacuum saturation-approximation~\cite{Lee:1973}.  Today, it is possible to calculate the \me{i} without appealing to this approximation and, therefore, without need for introducing the bag parameters.  
Nevertheless, because of their historical use, at a later stage in this project, we intend to combine results for \me{i} with our collaboration's results for $f_{B_q}$ to extract the bag parameters in a manner that accounts for correlations between the two calculations.
In this work
we present preliminary results for the matrix elements and refer the reader to Refs.~\cite{Bazavov:2011gy, newf}, and references therein, for values of $f_{B_q}$ calculated with similar fermion actions and gauge configurations.

In Sec.~\ref{sec-calc} we describe details of our calculation.
A discussion of the full error budget appears in Sec.~\ref{sec-error}.
In Sec.~\ref{sec-results} we report preliminary results,
and in Sec.~\ref{sec-summary} we summarize and discuss the remaining work to be done.

\section{Calculation}
\label{sec-calc}

In this section we review the generation of correlation function data, the extraction of matrix elements from fits to the data, and the renormalization and chiral-continuum extrapolation of the matrix elements.

\subsection{Generating Correlation Function Data}
\label{sec-gendata}

Correlation function data are generated using the MILC gauge configurations~\cite{Bazavov:2010} with \mbox{$2+1$} flavors of asqtad staggered~\cite{asqtad} sea quarks.
Bottom valence quarks are simulated using the Fermilab interpretation of the Sheikholeslami-Wohlert (clover) action~\cite{ElKhadra:1997} with hopping parameter $\kappa_b$ tuned to produce the observed $B_s$ meson mass~\cite{Bernard:2011}.
Light valence quarks are simulated with the asqtad action.  
We work in the meson rest frame. 

Table~\ref{tab-ens} lists the ensembles used to date and selected valence quark parameters.
For each of the $n_{\rm conf}$ configurations we average over $n_{\mathrm{src}}$ sources with temporal spacing $T/4$ and spatial distribution randomized to reduce correlations.
At this stage of the calculation, different subsets of data have been analyzed for the different operators.  The subsets used are noted in Table~\ref{tab-ens}.
\vspace{0.1in}
\bt{ c @{$\;\;\;$} c @{$\;\;\;$}  c @{$\;\;\;$}  c @{$\;\;\;$} c @{$\;\;\;\;$} c @{$\;\;\;$}  c @{$\;\;\;\;$}        c @{$\;$} c @{$\;$} c}                                 
$a$ [fm] & $(\frac{L}{a})^{\, 3} \times \frac{T}{a} $\vspace{0.05in} & $u_0$ & ($am_l$, $am_h$)  & $am_q$ & $\kappa_b$ & $n_{\rm conf} \times n_{\mathrm{src}}$ & $\langle{\cal O}_{1,2} \rangle$ & $\langle{\cal O}_3\rangle$ & $\langle{\cal O}_{4,5}\rangle$ \\  
\hline
& & & \vspace{-0.15in} \\
$0.12$   & $24^3\times$64   &  0.8678   & (0.005, 0.05)        & val$_{0.12}$  & 0.0901 & $2099 \times 4$	&  X	&  X	&  X \\ 
$0.12$   & $20^3\times$64   &  0.8678   & (0.007, 0.05)        & val$_{0.12}$ & 0.0901 & $2110 \times 4$ 	&  X	&  X	&  X \\ 
$0.12$   & $20^3\times$64   &  0.8677   & (0.010, 0.05)          & val$_{0.12}$ & 0.0901 & $2259 \times 4$   	&  X	&  X	&  X \\ 
$0.12$   & $20^3\times$64   &  0.8688   & (0.020, 0.05)          & val$_{0.12}$ & 0.0918 & $2052 \times 4$ 	&  X	&  X	&  X \\ 
 & & & & & & \vspace{-0.15in} \\
$0.09$   & $40^3\times$96   &  0.8799   & (0.0031, 0.031)    & val$_{0.09}$ & 0.0976 & $1015 \times 4$  	&  X	&  X	&  X \\ 
$0.09$   & $32^3\times$96   &  0.8781   & (0.00465, 0.031)  & val$_{0.09}$ & 0.0977& $984 \times 4$  		&  X	&  X	&  X \\ 
$0.09$   & $28^3\times$96   &  0.8782   & (0.0062, 0.031)    & val$_{0.09}$ & 0.0979 & $1931 \times 4$  	&  	&  	&  X \\ 
$0.09$   & $28^3\times$96   &  0.8788   & (0.0124, 0.031)    & val$_{0.09}$ & 0.0982 & $1996 \times 4$  	&  	&  	&  X \\ 
 & & & & & &  \vspace{-0.15in} \\         
$0.06$   & $64^3\times$144 &  0.88764 & (0.0018, 0.018)    & val$_{0.06}$ & 0.1052 & $828 \times 4$ 		&  X	&  	&   \\            
 & & & & & &   \vspace{-0.15in} \\ \hline
\et{Summary of the MILC ensembles and valence masses used in this analysis.  
Each lattice has an approximate spacing $a$, spatial extent $L$, and temporal extent $T$.  The tadpole improvement factor is determined from the expectation value of the average plaquette, $u_0=\langle \mathrm{plaquette}\rangle ^{1/4}$.  The light and strange sea-quark masses are $m_l$ and $m_h$, respectively.  The sets of valence masses used, $m_q$, generally run from $m_h/10$ to $m_h$, with values given by  
val$_{0.12}$ = \{0.005, 0.007, 0.01, 0.02, 0.03, 0.0349, 0.0415, 0.05\}, 
 val$_{0.09}$ = \{0.0031, 0.0047, 0.0062, 0.0093,  0.0124, 0.0261, 0.031\}, 
 and  val$_{0.06}$ = \{0.0018, 0.0025, 0.0036, 0.0054, 0.0072, 0.016, 0.0188\}.  The bottom quark hopping parameter is $\kappa_b$. 
An ``X'' marks ensembles used in the the analysis of each operator.}{tab-ens}

\begin{figure}[t]
 \vspace{0.0in}
  \begin{center}
  {\scalebox{1.3}{\includegraphics[angle=0,width=0.41\textwidth]{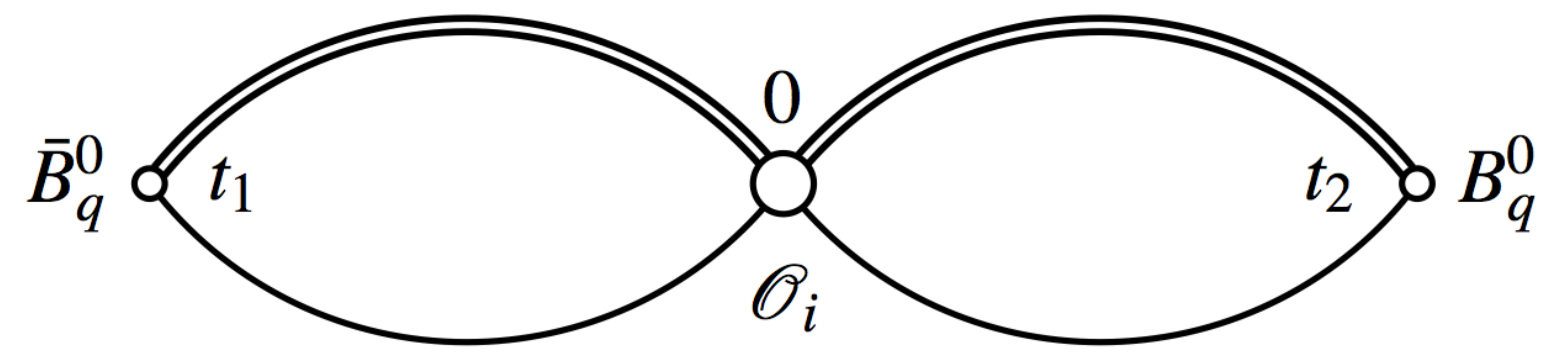}}}
    \vspace{0.1in}
    \caption[Three point correlation function setup.]{A $\bar{B}^0_q$ is created at rest at $t_1<0$.  At time $0$, via mixing operator ${\cal O}_i$, it oscillates into a $B^0_q$, which is subsequently annihilated at $t_2>0$.  The heavy (light) quark propagator is indicated by a double (single) line.  To extract the hadronic mixing matrix element we need the three-point correlation function depicted here and the two-point correlation functions corresponding to the propagation of the $\bar{B}_q^0$ from $t_1$ to 0 and of the $B_q^0$ from 0 to $t_2$.}
    \label{fig-setup}
    \vspace{-0.0in}
  \end{center}
\end{figure}

Data are generated as ensemble average estimates of the vacuum expectation values
\begin{eqnarray}
C^{\mathrm{2pt}}(t) &=& \sum_{{\bf x}} \Big\langle B_q^0(t,\, {\bf x})\ B_q^0(0,\, {\bf 0})^{\dagger} \Big\rangle
\label{eq-2pt} 
\end{eqnarray}
and   
\begin{eqnarray}
C_i^{\mathrm{3pt}}(t_1,t_2) &=& \sum_{{\bf x}_1,\, {\bf x}_2}\Big\langle B_q^0(t_2,\, {\bf x}_2)\ {\cal O}_i(0,\, {\bf 0})\ B_q^0(t_1,\, {\bf x}_1)\Big\rangle.
\label{eq-3pt}
\end{eqnarray}
The $B$ meson creation operator is given by
\begin{equation}
B^0_q(t,\, {\bf x})^\dagger = \sum_{{\bf y}} \bar{\psi}(t,\, {\bf y}) S({\bf x},{\bf y})\ \gamma_5\ q(t,\, {\bf x}),
\end{equation}
and the four-quark mixing operator is
\begin{equation}
\mathcal{O}_i(0,{\bf 0}) = \bar{\psi}(0,\, {\bf 0})\ \Gamma_i^{(1)} q(0,\, {\bf 0})\ \ \bar{\psi}(0,\, {\bf 0})\ \Gamma_i^{(2)} q(0,\, {\bf 0}).
\end{equation}
Propagators arising from the Wick contraction of naive light-quark fields $q$ are related to staggered quark propagators, generated in our simulation, following Ref.~\cite{Wingate:2003}.
The heavy-quark field $\psi$ is a four-component Dirac spinor, smeared in the $B$ meson creation operator to improve overlap with the ground state.  The smearing function $S({\bf x},{\bf y})$ is based on the quarkonium 1S wavefunction~\cite{DiPierro:2002}.  Heavy-quark fields in the mixing operator are rotated to remove leading order discretization errors~\cite{ElKhadra:1997}.  
The $\Gamma_i^{(1,2)}$ represent the spin structures appearing in the quark bilinears of Eq.~(\ref{eq:susybasis}).
The three-point correlation function setup is illustrated in Fig.~\ref{fig-setup}.  

Although parity conservation of QCD ensures the hadronic mixing matrix elements of $\tilde\mathcal{O}_{1,2,3}$ are physically equivalent to those of $\mathcal{O}_{1,2,3}$, the finite statistics estimates of Eq.~(\ref{eq-3pt}) need not generate data that are numerically identical.
We average these physically equivalent data to increase statistics associated with the determination of the matrix elements of $\mathcal{O}_{1,2,3}$.

\subsection{Fitting Correlation Function Data}
\label{sec-fitdata}
We model two- and three-point correlation function data by projecting the vacuum expectation values of Eqs.~(\ref{eq-2pt}) and (\ref{eq-3pt}) onto bases of $B^0_q$ and $\bar{B}^0_q$ energy eigenstates,
\begin{eqnarray}
C^{\mathrm{2pt}}(t) &=& \sum_{n=0}^{N^{\mathrm{2pt}}-1} Z^2_n\ (-1)^{n(t+1)} (e^{-E_n t} + e^{-E_n(T-t)}), \label{eq-2ptfitfn} \\
& & \nonumber \\
C^{\mathrm{3pt}}_i(t_1,t_2) &=& \sum_{n,\, m=0}^{N^{\mathrm{3pt}}-1} \frac{Z_n Z_m}{2}\ \left( \frac{\langle (B_q^0)_n | {\cal O}_i |(\bar{B}_q^0)_m\rangle}{\sqrt{E_n E_m}}\right)\ (-1)^{m(t_1+1)+n(t_2+1)} e^{-E_nt_2}\ e^{-E_mt_1},
\label{eq-3ptfitfn}
\end{eqnarray}
where oscillating, opposite-parity state contributions~\cite{Wingate:2003} are incorporated.  The effects of periodic boundary conditions are negligible in fitted three-point data and are omitted from Eq.~(\ref{eq-3ptfitfn}).

We extract the ground state quantity $\langle B_q | {\cal O}_i | \bar{B}_q\rangle / M_{B_q}$ by simultaneously fitting the two- and three-point correlation function data to Eqs.~(\ref{eq-2ptfitfn}) and (\ref{eq-3ptfitfn}).  
In principle the sums run over infinitely many radial excitations of the pseudoscalar $B$ meson, in practice we must fit with finite $N^{\mathrm{2,\, 3pt}}$.
We also restrict data included in the fit $(t_{\mathrm{min}}\leq t \leq t_{\mathrm{max}})$.  Fits using data at short times must account for excited state contributions by including adequate numbers of states.  Despite added difficulty, the relatively clean signal in the data at short times makes it desirable to include them in the fit.  The use of a Bayesian fitting routine~\cite{Lepage:2001ym,Lepage:2008} allows us to deal with the difficulty of fitting numerous states and incorporate data at short times.  Increasing $t_{\mathrm{max}}$ has the benefit of utilizing more data.  However, this also introduces an increasing level of noise and can lead to a poorly determined covariance matrix.

From plots of correlation function data, scaled to remove leading exponential decay, we determine a range of potential $t^{\mathrm{2,\, 3pt}}_{\mathrm{min}}$ and $t^{\mathrm{2,\, 3pt}}_{\mathrm{max}}$ values to study.\footnote{For $C^{\mathrm{3pt}}_i(t_1,t_2)$, 
we could choose distinct $t_{1, \rm min}$ and $t_{2, \rm min}$ and distinct $t_{1, \rm max}$ and $t_{2, \rm max}$.  However, an explicit $t_1 \leftrightarrow t_2$ symmetry of data generated by Eq.~(\ref{eq-3pt}) prompts the simplifications:  $t^{\mathrm{3pt}}_{\mathrm{min}}\equiv t_{1,\, \mathrm{min}}=t_{2,\, \mathrm{min}}$ and $t^{\mathrm{3pt}}_{\mathrm{max}}\equiv t_{1,\, \mathrm{max}}=t_{2,\, \mathrm{max}}$.} 
The range of potential $t^{\mathrm{2,\, 3pt}}_{\mathrm{max}}$ values is determined by the onset of significant noise.  For $t^{\mathrm{2,\, 3pt}}_{\mathrm{min}}$ we consider from $t^{\mathrm{2,\, 3pt}}_{\mathrm{min}} = 2$ until excited state contributions have significantly decreased.  
We perform fits with $N^{\mathrm{2,\, 3pt}}=2,4$, and (sometimes) 6.\footnote{We use equal numbers of parity-even and parity-odd states in our fits.}  
Starting with the two-point correlation functions, we fit each combination of $N^{\mathrm{2pt}}$, $t^{\mathrm{2pt}}_{\mathrm{min}}$, and $t^{\mathrm{2pt}}_{\mathrm{max}}$ and select representative fits at each $N^{\rm 2pt}$ from a common plateau of fit results for $Z_0$ and $E_0$ versus $t^{\mathrm{2pt}}_{\mathrm{min}}$ and $t^{\mathrm{2pt}}_{\mathrm{max}}$, thereby ensuring stability with respect to $N^{\mathrm{2pt}}$, $t^{\mathrm{2pt}}_{\mathrm{min}}$, and $t^{\mathrm{2pt}}_{\mathrm{max}}$. 
Then, fixing $t^{\mathrm{2pt}}_{\mathrm{min}}$ and $t^{\mathrm{2pt}}_{\mathrm{max}}$ separately for each $N^{\rm 2pt}$,
we repeat the procedure for a simultaneous fit of the two- and three-point correlation function data.  For these fits, we fix $N^{\mathrm{2pt}}=N^{\mathrm{3pt}}$ and determine the optimum $N^{\mathrm{3pt}}$, $t^{\mathrm{3pt}}_{\mathrm{min}}$, and $t^{\mathrm{3pt}}_{\mathrm{max}}$.
This systematic procedure, outlined in more detail in~\cite{Bouchard:2011, Bouchard:2010}, allows us to obtain fit results that are stable in the number of states and data included in the fit.  We find the fit results are largely insensitive to changes in $t^{\mathrm{2,\, 3pt}}_{\mathrm{max}}$.

\subsection{Renormalization}	
\label{sec-renormalization}
Lattice results for the hadronic mixing matrix elements are combined with perturbatively calculated Wilson coefficients, {\it i.e.} $C_i$ and $\tilde{C}_i$ of Eq.~(\ref{eq:hamiltonian}), to obtain physical quantities.  
Therefore, the lattice and continuum results must be brought to the same scheme and scale.  
We use the mass of the bottom quark to set the scale and match to the continuum $\overline{\rm MS}-$NDR scheme.  At one loop the matrix elements mix under renormalization,
\begin{equation}
\langle B | \mathcal{O}_i|\bar{B}\rangle^{\rm \overline{MS}-NDR}(m_b) =  \sum_{j=1}^5 \big[ \delta_{ij}+\alpha_s(q^*) \zeta_{ij}(am_b)\big] \langle B | \mathcal{O}_j|\bar{B}\rangle^{\rm lat} ,
\label{eq-renorm}
\end{equation}
where the matching coefficients $\zeta_{ij}$ give the difference between the one-loop lattice and continuum renormalizations~\cite{Gamiz:inprep}.   
We evaluate $\alpha_s(q^*)$, as in~\cite{Mason:2005}, in the ``V scheme''~\cite{Lepage:1993} at the scale $q^*=2/a$. $\overline{\rm MS}$-NDR is not sufficient to define the matrix elements of operators ${\mathcal O}_2$ and ${\mathcal O}_3$ in dimensional regularization; one must also choose a set of ``evanescent'' operators. The renormalization coefficients $\zeta_{2j}$ and $\zeta_{3j}$ must be calculated accordingly. 
Below, we present results for two schemes, which we label ``BBGLN''~\cite{Beneke:1998sy} and ``BJU''~\cite{Buras:2001ra}.
The one-loop conversion between the two schemes was first given in Ref.~\cite{Becirevic:2001xt}.
A more general transformation of the two-loop anomalous dimension matrix can be found in \cite{Gorbahn:2009pp}. 
For convenience, we define the shorthand  
$\langle {\mathcal O}_i \rangle \equiv \langle B | {\mathcal O}_i|\bar{B}\rangle^{\rm \overline{MS}-NDR}(m_b)$,   
for the matrix elements after extrapolating the renormalized results to the continuum.

\subsection{Chiral-Continuum Extrapolation}
\label{sec-chipt}

The continuum, next-to-leading order (NLO), chiral perturbation theory for hadronic mixing matrix elements~\cite{Detmold:2007}, modified~\cite{oldBBbar} for the rooted, staggered, light-quark actions (rS$\chi$PT), guides a simultaneous extrapolation to physical quark mass and the continuum.  
The rS$\chi$PT expression is fit to the renormalized matrix elements of Eq.~(\ref{eq-renorm}) to determine the low energy constants.  
Effects from heavy-light meson flavor- and hyperfine-splittings~\cite{Bazavov:2011gy} and finite volume (expected to be $< 1\%$) are not explicitly incorporated in the chiral extrapolations presented here, but will be in the final stage of analysis.  Chiral fits are performed using a Bayesian fitting routine~\cite{Lepage:2001ym,Lepage:2008} with priors constraining each fit parameter.  
Additional details 
will be given in future publications.  

A chiral expansion including only NLO terms fits the data  reasonably well when data included in the fit are limited to light valence masses ({\it i.e.}, values of $m_q$ such that the pseudoscalar meson mass of Eq.~(\ref{eq:ps}) satisfies $(M_{qq}\ r_1)^2 \lesssim 1$).\footnote{For reference, at the strange valence mass, $(M_{qq}\ r_1)^2 \sim 1.3$.}
To include data at all simulated valence masses we find it necessary to include NNLO analytic terms.  
There are seven potential NNLO analytic terms~\cite{Evans:thesis}
\begin{equation}
a^2 M^2_{qq},\ \ a^2 (2M^2_{ll}+M^2_{hh}),\ \ M^2_{qq}(2M^2_{ll} + M^2_{hh}),\ \ M_{qq}^4,\ \ (2M^2_{ll} + M^2_{hh})^2,\ \ a^4,\ \ (2M_{ll}^4 + M_{hh}^4),
\label{eq-NNLOterms}
\end{equation}
where the (squared) pseudoscalar meson mass is given by 
\begin{equation}	\label{eq:ps}
M^2_{ij} = B_0(m_i + m_j)
\end{equation}
and $B_0$ is a low-energy constant from $\chi$PT.  Each of these terms comes with a coefficient to be determined by the fit.  At a minimum, we find it necessary to include terms proportional to $a^2 M^2_{qq}$ and $M^2_{qq}(2M^2_{ll} + M^2_{hh})$ in order to achieve suitable fits to data at all valence masses.

\begin{figure}[b]
	\begin{center}
		\vspace{-0.3in}
		\includegraphics[width=0.87 \textwidth]{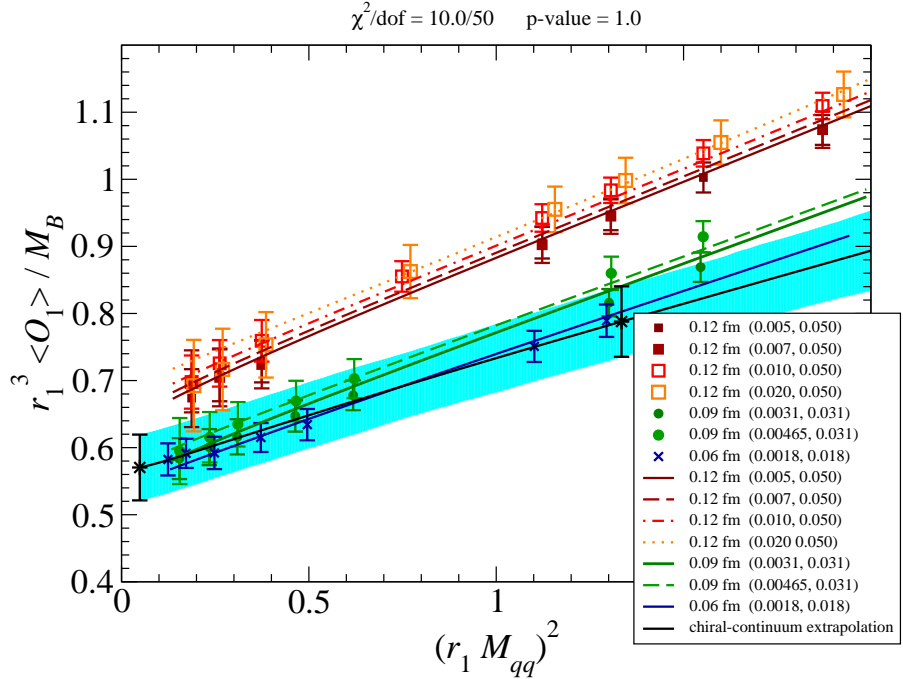}
		\vspace{-0.2in}
		\caption{$\langle \mathcal{O}_1 \rangle$:  chiral-continuum extrapolation data and fits.  (Red/orange) squares are 0.12 fm data, (green) circles are 0.09 fm data, and (blue) $\times$'s are 0.06 fm data.  The extrapolation/interpolation is the black line with a shaded (turquoise) error band.  Black bursts mark the point of extrapolation (interpolation) for the $B_d$ ($B_s$) meson.}
		\label{fig-chiptO1}
	\end{center}
\end{figure}

\begin{figure}[h!]
	\begin{center}
		\vspace{-0.51in}
		\includegraphics[width=0.87 \textwidth]{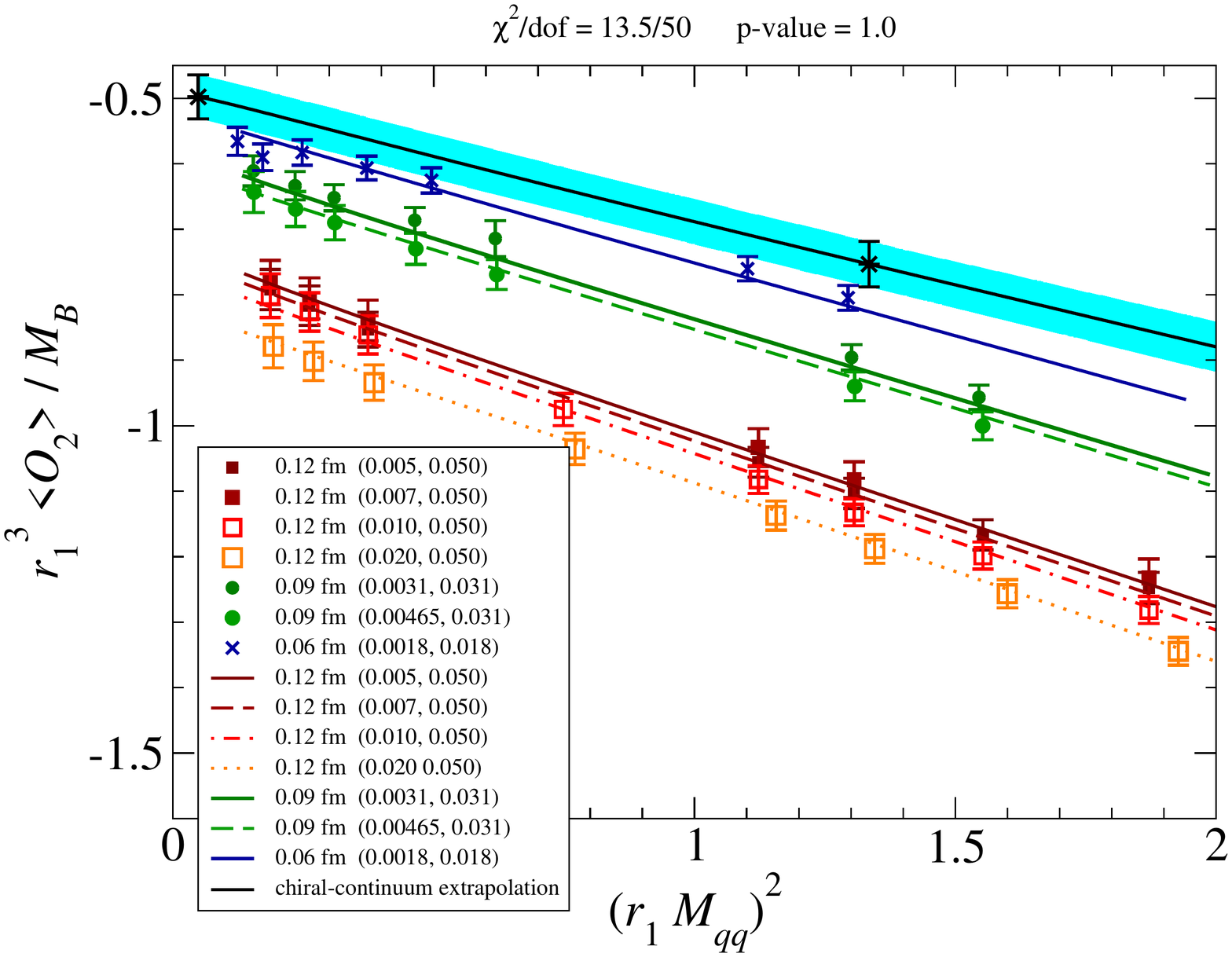}
		\vspace{-0.24in}
		\caption{$\langle \mathcal{O}_2 \rangle$:  chiral-continuum extrapolation data and fits in the BBGLN scheme described in Sec.~\protect\ref{sec-renormalization}.  (Red/orange) squares are 0.12 fm data, (green) circles are 0.09 fm data, and (blue) $\times$'s are 0.06 fm data.  The extrapolation/interpolation is the black line with a shaded (turquoise) error band.  Black bursts mark the point of extrapolation (interpolation) for the $B_d$ ($B_s$) meson.}
		\label{fig-chiptO2}
	\end{center}
\end{figure}

\begin{figure}[h!]
	\begin{center}
		\vspace{-0.35in}
		\includegraphics[width=0.87 \textwidth]{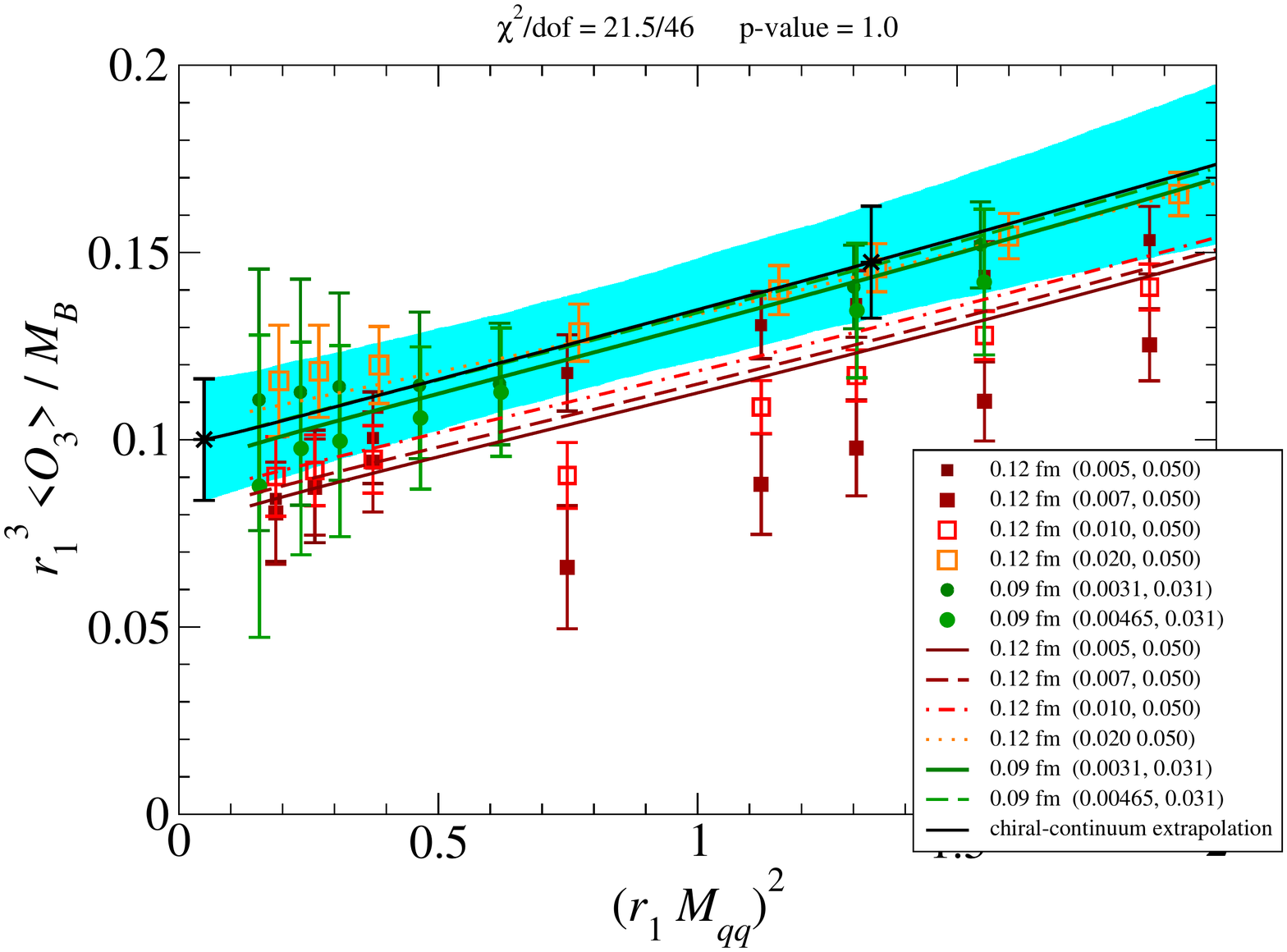}
		\vspace{-0.24in}
		\caption{$\langle \mathcal{O}_3 \rangle$:  chiral-continuum extrapolation data and fits in the BBGLN scheme described in Sec.~\protect\ref{sec-renormalization}.  (Red/orange) squares are 0.12 fm data and (green) circles are 0.09 fm data.  The extrapolation/interpolation is the black line with a shaded (turquoise) error band.  Black bursts mark the point of extrapolation (interpolation) for the $B_d$ ($B_s$) meson.}
		\vspace{-0.6in}
		\label{fig-chiptO3}
	\end{center}
\end{figure}

\begin{figure}[h!]
 \vspace{-0.1in}
  \begin{center}
  \includegraphics[width=0.9 \textwidth]{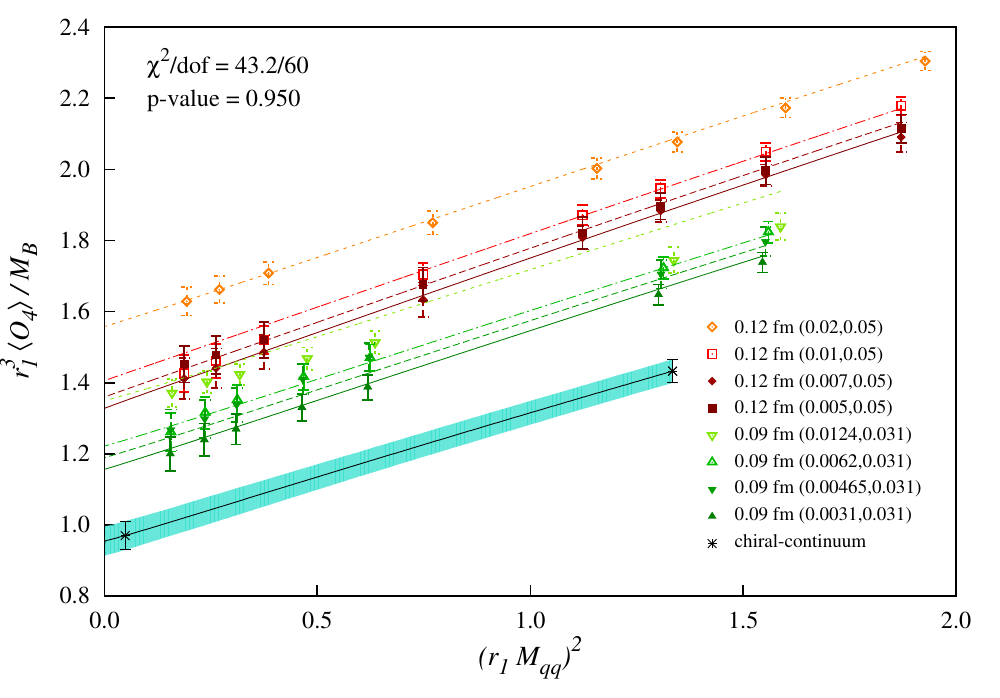}
    \vspace{-0.2in}
    \caption{$\langle \mathcal{O}_4 \rangle$:  chiral-continuum extrapolation data and fits.  (Red/orange) squares are 0.12 fm data, (green) triangles are 0.09 fm data.  The extrapolation/interpolation is the black line with a shaded (turquoise) error band.  Black bursts mark the point of extrapolation (interpolation) for the $B_d$ ($B_s$) meson.}
    \label{fig-chiptO4}
  \end{center}
\end{figure}

\begin{figure}[h!]
 \vspace{-0.1in}
  \begin{center}
  \includegraphics[width=0.9 \textwidth]{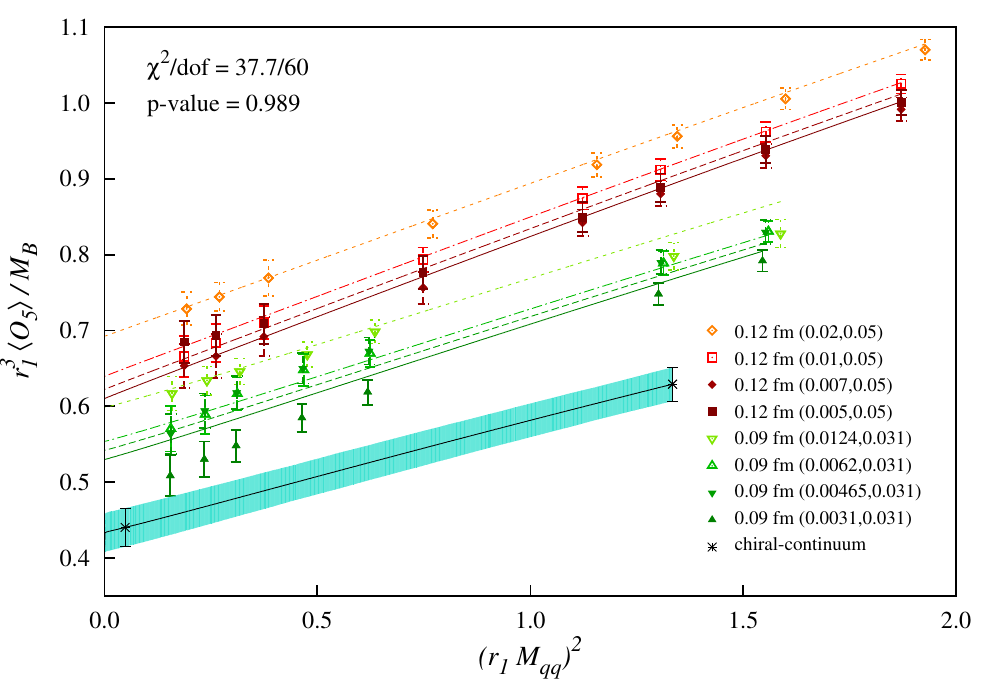}
    \vspace{-0.2in}
    \caption{$\langle \mathcal{O}_5 \rangle$:  chiral-continuum extrapolation data and fits.  (Red/orange) squares are 0.12 fm data, (green) triangles are 0.09 fm data.  The extrapolation/interpolation is the black line with a shaded (turquoise) error band.  Black bursts mark the point of extrapolation (interpolation) for the $B_d$ ($B_s$) meson.}
    \label{fig-chiptO5}
  \end{center}
\end{figure}

Reported central values are obtained from those fits including all seven NNLO terms.  The simultaneous chiral-continuum extrapolations of $\langle {\mathcal O}_i\rangle / M_B$ are shown in Figs.~\ref{fig-chiptO1}, \ref{fig-chiptO2}, \ref{fig-chiptO3}, \ref{fig-chiptO4}, and \ref{fig-chiptO5}.
The p-values of these fits are very near one.  This is not the case for NLO fits or fits including various subsets of the possible NNLO analytic terms of Eq.~(\ref{eq-NNLOterms}).  Inflated p-values, at least partially, result from the fact that we use the Bayesian augmented $\chi^2$ and degrees of freedom.  In addition, the size of our statistical sample may be limiting our ability to resolve small eigenvalues of the correlation matrix. We are studying this issue.
%

\clearpage

\section{Errors}
\label{sec-error}

The error budget given in this work is complete in that it accounts for all known, significant sources of error in our lattice-QCD calculation. 
We have split the list of errors into two tables.  
Table~\ref{tab-stat} lists the statistical and chiral-continuum errors, which are determined directly from the analysis presented here, for each $\langle{\cal O}_i\rangle$ individually.
Errors estimated from our collaboration's previous works~\cite{Bazavov:2011gy, Evans:2009} are listed in Table~\ref{tab-syst} and are taken (for now) to be the same for all $\langle{\cal O}_i\rangle$.
 Errors associated with omitting charm and heavier sea-quark contributions, isospin breaking, and electromagnetic effects are expected to be insignificant in comparison to the errors in Tables~\ref{tab-stat} and~\ref{tab-syst}. 
 Further study, and an ongoing analysis of $a\approx0.06$ fm and $a\approx0.045$ fm data, should lead to a reduction of our leading sources of error in Tables~\ref{tab-stat}~and~\ref{tab-syst}.
 
\renewcommand*\arraystretch{1.2}
\begin{table}[t!]
\begin{center}
\begin{tabular}{ll ccccc  @{$\qquad$} cc}
	\hline\hline
	                                      	& Source of Error [\%]        	& \shortop1 	& \shortop2 	& \shortop3 	& \shortop4 	& \shortop5	  \\  
	\hline
	\multirow{2}{*}{$B^0_d$} 	& statistical                            	&   8.6            	&    6.8           	&   16      		& 4.3              	& 5.5 	  \\              
	                                        	& chiral-continuum systematic 	&   12           	&    11           	&   3.3          	& 0.2             	& 4.4	 \\            
	\hline
	\multirow{2}{*}{$B^0_s$} 	& statistical                            	&   6.7            	&   4.6             	&   10             	& 2.5              	& 3.4    \\
	                                        	& chiral-continuum systematic 	&   1.8           	&   6.6            	&   4.5              	& 1.6              	& 3.7  \\
	\hline\hline
\end{tabular}
\caption{Statistical and chiral-continuum systematic errors for the hadronic mixing matrix elements are determined from the data by methods outlined in Sec.~\protect\ref{sec-error}.  Here we show results from the use of the BBGLN  evanescent operators in the renormalization, as described in Sec.~\protect\ref{sec-renormalization}.
To the precision reported here, errors in the BJU scheme differ only for the chiral-continuum systematic error for $B^0_s$ mixing via $\shortop2$, where we found the error to be 6.4\%.}
\label{tab-stat}
\vspace{0.3in}
\begin{tabular}{lc}
	\hline\hline
	Source of Error [\%]                        & $\langle{\cal O}_i\rangle$ \\  
	\hline
	scale ($r_1$)                                   & 3                           \\
	$\kappa_b$ tuning                           & 4                          \\	
	light-quark masses                          & 1                               \\
	heavy-quark discretization               & 4                          \\
	one-loop matching                           &  8                         \\
	finite-volume effects                         & 1                               \\
	\hline
	subtotal                                             & 10                        \\
	\hline\hline
\end{tabular}
\end{center}
\vspace{-0.15in} 
\caption{Estimates of errors based on our collaboration's previous works~\cite{Bazavov:2011gy, Evans:2009}.  We anticipate significant reduction in the leading sources of error once all ensembles at $a \approx 0.06$ fm and at $a \approx 0.45$ fm have been fully incorporated into the analysis.}
\label{tab-syst}
\end{table}

Statistical errors are obtained by propagating a distribution of 600 bootstrap ensembles of correlation function data ({\it i.e.}, Eqs.~(\ref{eq-2ptfitfn} and \ref{eq-3ptfitfn})) through the correlation function and rS$\chi$PT fitting procedures, providing non-parametric error estimates that account for correlations among the data.  This error also includes uncertainty associated with chiral fit parameters $f_\pi$ and $g_{B^*B\pi}$ by setting prior widths based on parameter uncertainties.

The systematic error associated with the chiral-continuum series truncation and extrapolation is obtained by varying the rS$\chi$PT fit ansatz.  For both the $B^0_d$ and $B^0_s$ matrix elements, the addition of the two minimum required NNLO analytic terms of Eq.~(\ref{eq-NNLOterms}) is necessary, and sufficient, to achieve good fits to all valence mass data that are stable against the addition of remaining NNLO terms.  We estimate the systematic error to be the shift in fit central values between fits with the two required NNLO analytic terms and those with all NNLO analytic terms.

The remaining errors, quoted in Table~\ref{tab-syst}, are conservative estimates based on previous collaboration publications.  
The errors associated with scale-setting and $\kappa_b$ tuning are estimated from Ref.~\cite{Bazavov:2011gy}.  Entries for light-quark mass, heavy-quark discretization, and one-loop matching errors are estimated from Ref.~\cite{Evans:2009}.
Finite volume error estimates are based on Ref.~\cite{Evans:2009} and inflated to accommodate an expected increase~\cite{Arndt:2004} from hyperfine- and flavor-splitting effects.

\section{Results}	
\label{sec-results}
Results from our chiral-continuum extrapolations are in $r_1$ units.  We convert to GeV using the lattice determination $r_1=0.3117(22)$~fm~\cite{Bazavov:2011gy} and the conversion factor $1\ \textrm{GeV\ fm} = 5.0677312(1)$.  
In lieu of the matrix elements, in Table~\ref{tab-results} we report the phenomenologically more useful quantity $f^2_{B_q}\ B^{(i)}_{B_q}$.
Results are given in the continuum $\overline{\rm MS}-$NDR scheme at scale $m_b$ using both the BBGLN~\cite{Beneke:1998sy} and BJU~\cite{Buras:2001ra} choices of evanescent operators.
Applying a heavy-quark relation between the matrix elements of ${\mathcal O}_1$, ${\mathcal O}_2$, and ${\mathcal O}_3$, the conversion~\cite{Becirevic:2001xt} for bag parameters can be expressed as
\begin{eqnarray}
	B_{B_q}^{(2)BJU} &=& B_{B_q}^{(2)BBGLN} + \frac{\alpha_s}{4\pi} \left(
		\frac{4}{15}B_{B_q}^{(1)} - 4 B_{B_q}^{(2)} \right), 	\label{eq:B2} \\
	B_{B_q}^{(3)BJU} &=& B_{B_q}^{(3)BBGLN} + \frac{\alpha_s}{4\pi} \left(
		-\frac{4}{3}B_{B_q}^{(1)} + \frac{4}{3}B_{B_q}^{(3)} \right).	\label{eq:B3}
\end{eqnarray}
For the results given in Table~\ref{tab-results}, we bring the matrix elements to the BBGLN or BJU scheme before the chiral-continuum extrapolation.
Converting \emph{a~posteriori} with Eqs.~(\ref{eq:B2}) and~(\ref{eq:B3}) yields values in very good agreement.

The scheme dependence of the $f^2_{B_q}\ B^{(i)}_{B_q}$ from Table~\ref{tab-results} matches the 
corresponding scheme dependence in the Wilson coefficients with which they will be used.
In particular, one must use the BBGLN or BJU values for $\shortop2$ and $\shortop3$, according to the choice of evanescent operator scheme used in the  Wilson coefficient calculation. 
As mentioned in Sec.~\ref{sec-renormalization},  $\shortop1$, $\shortop4$, and $\shortop5$ are independendent of this choice.

Errors are obtained by adding in quadrature the appropriate errors from Table~\ref{tab-stat} and the subtotal of Table~\ref{tab-syst}.
As shown in Table~\ref{tab-stat} and illustrated in Fig.~\ref{fig-chiptO3}, the matrix element \me3 suffers from much larger errors than the others and is more difficult to analyze.  
We are optimistic that the addition of data at finer lattice spacings, and refinements in analysis techniques, will lead to improvements.  
\begin{table}[t!]
	\vspace{0.2in}
	\begin{center}
		\begin{tabular}{c   r@{.}l    r@{.}l               r@{.}l  r@{.}l  } 
		\hline \hline            
		 						& \mc{4}{c}{$B^0_d$}						& \mc{4}{c}{$B^0_s$}	\\
		$[{\rm GeV}^2]$			& \mc{2}{c}{BBGLN}		& \mc{2}{c}{BJU}		& \mc{2}{c}{BBGLN}		& \mc{2}{c}{BJU}	\\
		\hline
		 $f_{B_q}^2 B_{B_q}^{(1)}$	&  \mc{4}{c}{0.0411(75)}						& \mc{4}{c}{0.0559(68)}	\\ [0.5em]
		 $f_{B_q}^2 B_{B_q}^{(2)}$	& 0&0574(92)			& 0&0538(87)			& 0&086(11) 	&  0&080(10)\\ [0.5em]
		 $f_{B_q}^2 B_{B_q}^{(3)}$	& 0&058(11)			&  0&058(11)			& 0&084(13)	&  0&084(13)	 \\ [0.5em]
		 $f_{B_q}^2 B_{B_q}^{(4)}$	& \mc{4}{c}{0.093(10)} 						& \mc{4}{c}{0.135(15)}	 \\ [0.5em]
		 $f_{B_q}^2 B_{B_q}^{(5)}$	& \mc{4}{c}{0.127(15)} 						& \mc{4}{c}{0.178(20)}\\ [0.5em]
		 \hline \hline
		\end{tabular}
	\end{center}
	\vspace{-.15in}
	\caption{Preliminary results for $f_{B_q}^2 B_{B_q}^{(i)} =  \langle{\cal O}_i\rangle\, /\, \mathfrak{c_i} M^2_{B_q} $, evaluated in the continuum $\overline{\rm MS}-$NDR scheme at scale $m_b$. 
	``BBGLN'' and ``BJU'' denote the choice of evanescent operators in the renormalization schemes of Refs.~\cite{Beneke:1998sy,Buras:2001ra}, respectively. For $i=1,4,5$, the schemes are the same.    
	Although the chiral-continuum extrapolation was done separately for both schemes, its output for $f_{B_q}^2 B_{B_q}^{(3)}$ is the same for the number of digits shown.
	The $B^0_q$ meson masses, needed in combination with the quantity extracted in Eq.~(\protect\ref{eq-3ptfitfn}), 
	are known to $\lesssim 0.01\%$~\protect\cite{Nakamura:2010zzi} and do not contribute significantly to the error.
	}
	\label{tab-results}	
\end{table}

\section{Summary and Outlook}
\label{sec-summary}

We report on our ongoing, $2+1$ flavor, lattice-QCD calculation of the hadronic contribution to $B^0$ and $B^0_s$ mixing.  Table~\ref{tab-results} gives preliminary results for the hadronic mixing matrix elements for a basis of four-quark, dimension-six, $\Delta B=2$ mixing operators that spans the space of all possible hadronic contributions in the Standard Model and beyond.
At the current stage of analysis our errors are competitive with published Standard Model matrix element results.
For beyond the Standard Model matrix elements, this is the first unquenched calculation and the first update in ten years.

Additional data at finer lattice spacings, $a\approx 0.06$ fm and $a \approx 0.045$ fm, is currently being analyzed.  
We anticipate that its inclusion, and a more rigorous analysis of the errors of Table~\ref{tab-syst}, will result in a significant reduction in the error budget.
The discretization and one-loop matching errors of Table~\ref{tab-syst} are based on $a\approx 0.09$~fm data, and will decrease when re-evaluated for $a~\approx 0.045$~fm.  Recent data runs will allow for a new analysis of $\kappa_b$ tuning errors as well.
We are finalizing the calculation of the one-loop perturbative matching coefficients of Sec.~\ref{sec-renormalization} and plan to incorporate finite volume effects and hyperfine- and flavor-splittings in the chiral-continuum extrapolation.

\section*{Acknowledgments}
\label{sec-ack}
Computations for this work were carried out with resources provided by the USQCD Collaboration, the Argonne
Leadership Computing Facility, the National Energy Research Scientific Computing Center, and the Los Alamos National Laboratory, which are funded by the Office of Science of the U.S. Department of Energy; and with resources provided by the National Center for Supercomputer Applications, the National Institute for Computational Science, the Pittsburgh Supercomputer Center, the San Diego
Supercomputer Center, and the Texas Advanced Computing Center, which are funded through the National Science
Foundation Teragrid/XSEDE Program.  
This work was supported in part by the U.S. Department of Energy under Grants No. DE-FG02-91ER40677 (A.X.E., E.D.F., and C.M.B) and No. DE-FG02-91ER40628 (C.B.); and by the Science and Technology Facilities Council and the Scottish Universities Physics Alliance (J.L.). 
C.M.B. was supported in part by a Fermilab Fellowship in Theoretical Physics.  Fermilab is operated by Fermi Research Alliance, LLC, under Contract No.~DE-AC02-07CH11359 with the United States Department of Energy.  This manuscript has been co-authored by employees of Brookhaven Science Associates, LLC, under Contract No. DE-AC02-98CH10886 with the U.S. Department of Energy.



\end{document}